\documentclass[pra,showpacs,superscriptaddress]{revtex4}

\usepackage{bm}

\input{epsf}

\begin{document}
\title{A multi-photon Stokes-parameter invariant for entangled states\\}

\author{Gregg Jaeger}

\affiliation {Department of Electrical and Computer Engineering \\
Boston University, Boston MA 02215}

\author{Mihail Teodorescu-Frumosu}

\affiliation {Department of Mathematics \\
Boston University, Boston MA 02215}

\author{Alexander Sergienko}

\affiliation {Department of Electrical and Computer Engineering \\
Boston University, Boston MA 02215}

\affiliation {Department of Physics\\
Boston University, Boston MA 02215}

\author{Bahaa~E.~A.~Saleh}

\affiliation {Department of Electrical and Computer Engineering \\
Boston University, Boston MA 02215}

\author{Malvin~C.~Teich}

\affiliation {Department of Electrical and Computer Engineering \\
Boston University, Boston MA 02215}

\affiliation {Department of Physics\\
Boston University, Boston MA 02215}

\date{\today}

\renewcommand\baselinestretch{2}\small\normalsize
\begin{abstract}

We consider the Minkowskian norm of the $n$-photon Stokes tensor,
a scalar invariant under the group realized by the transformations
of stochastic local quantum operations and classical
communications (SLOCC). This invariant is offered as a candidate
entanglement measure for $n$-qubit states and discussed in
relation to measures of quantum state entanglement for certain
important classes of two-qubit and three-qubit systems. This
invariant can be directly estimated via a quantum network,
obviating the need to perform laborious quantum state tomography.
We also show that this invariant directly captures the extent of
entanglement purification due to SLOCC filters.

\end{abstract}

\pacs{03.65.Ta,03.67.-a,42.79.Ta} \maketitle

\vfill\eject

Entangled states of photon polarization have been of ongoing
interest for their role in probing fundamental aspects of quantum
theory, in the coding and manipulation of quantum information and
in practical applications of quantum interferometry. In both
classical and quantum optics, Stokes parameters have proven
intuitive and practical tools for characterizing polarization
states of light. Here, we examine a group-invariant scalar measure
on the space of generalized Stokes parameters. We show that this
norm, which is an invariant under transformations of stochastic
local quantum operations and classical communications (SLOCC
\cite{18}) on $n$ qubits, quantifies entanglement for certain
important classes of two-qubit and three-qubit systems, and
potentially for similar classes of $n>3$ qubits. Our results for
several photon states, together with its mathematical properties
for all values of $n$, recommend this scalar as a candidate
measure of total entanglement for multi-particle pure states in
general. This invariant has the valuable property of being
directly estimable via a quantum network, in principle obviating
the need to perform quantum state tomography, an increasingly
laborious task as the number of particles increases, to determine
the density matrix first in order to find the degree of
entanglement. The invariant allows one to immediately identify the
SLOCC filtering transformations as entanglement purifiers and
directly captures the amount of purification achieved by these
filters or any other process.

\section{Definitions and the general case}

In classical optics, the four Stokes parameters, $S_\mu$ where
$\mu = 0, 1, 2, 3$, are known to form a four-vector under the
O$_0(1,3)$ group of transformations \cite{4,45,5,6}. These four
parameters characterize the time-averaged electric field intensity
and the distribution of polarization among three orthogonal
polarization directions in the Poincar\'e sphere. The associated
invariant length of the Stokes four-vector is $S^2\equiv S_0^2
-S_1^2 - S_2^2 - S_3^2$~\cite{6}. These transformations can be
represented by an ordinary rotation, followed by a hyperbolic
rotation, followed by another ordinary rotation \cite{Sternberg}.
As a practical example, we note that the angles of the ordinary
polarization rotations may parametrize the effect of birefringence
during light propagation in optical fiber; those of the hyperbolic
polarization rotations may parametrize the effect of dichroism in
fiber \cite{5}.

In the quantum case, the Stokes parameter representation of the
single-photon ensemble is formally similar to that of classical
polarization optics, and will similarly be seen here to form a
four-vector. To address the multiple-photon case, we make use of
$n$-photon generalized Stokes parameters (see, for example,
\cite{2}),
\begin{equation}\label{Stensor}
  S_{i_1...i_n}= \mathrm{Tr} (\rho\ \sigma_{i_1}\otimes
  ...\otimes\sigma_{i_n})\ , \ \ i_1, ...,i_n=0,1,2,3\ ,
\end{equation}

\noindent where $\sigma_\mu^2=1$, $\mu = 0, 1, 2, 3$, are the
three Pauli matrices together with the identity
$\sigma_0=I_{2\times 2}$, and ${1\over 2}\sigma_\mu\sigma_\nu
=\delta_{\mu\nu}$. These parameters form a full set of $n$-photon
generalized Stokes tensors $\{S_{i_1,...,i_n}\}$ that can be used
to describe coherence and entanglement properties of photon-number
states. The $n$-photon polarization density matrix can also be
conveniently written in terms of these generalized Stokes
parameters:

\begin{equation}\label{DensityN}
  \rho={1\over 2^n}\sum_{i_1,...,i_n=0}^3 S_{i_1...i_n}
  \sigma_{i_1}\otimes...\otimes\sigma_{i_n}\ ,\ \
  \ i_1, ...,i_n=0,1,2,3\ .
\end{equation}

Under SLOCC, the initial system density matrix undergoes
transformations of the group SL(2,C), while the multi-photon
Stokes parameters similarly undergo transformations of the group
O$_o(1,3)$, which we notate $S_{i_1... i_n}\rightarrow S'_{i_1...
i_n}$. The unitary subgroup of [SU(2)] transformations of the
two-qubit and three-qubit density matrices have been carefully
studied (see, for example, Ref. \cite{SM}). These correspond to
subgroup of ordinary [SO(3)] rotations of the quantum Stokes
tensor. The set of non-unitary [SL(2,C)$\setminus$SU(2)]
transformations of the density matrix have largely been overlooked
in the investigation of entanglement (with a few notable
exceptions \cite{G96,HHH,Vetal}). For the tensor of Stokes
parameters, these latter transformations [O$_0(1,3)\setminus$
SO(3)] involve hyperbolic rotation, corresponding physically to
polarization-dependent loss and intensity reduction:
$S'_{0...0}<1$. For these transformations, the Stokes parameters
$S'_{i   _1... i_n}$ must be renormalized due to the associated
removal of a portion of the original quantum ensemble, resulting
in the renormalized, physical values $S''_{i_1 ...i_n}=S'_{i_1 ...
i_n}/S'_{0...0}$. This will later allow us to identify a class of
filters that purify entanglement.

For each finite number $n$ of entangled photons, let us examine
the O$_0(1,3)$-group-invariant length, namely the Minkowskian
squared-norm of the Stokes tensor $\{S_{i_1... i_n}\}$, which we
refer to as the ``Stokes scalar." For reasons of convenience, we
choose to normalize this quantity by the factor $2^{-n}$:
\begin{eqnarray}\label{Invar}
  S_{(n)}^2&\equiv &{1\over 2^n}\bigg\{(S_{0 ...0})^2
-\sum_{k=1}^n\sum_{i_k=1}^3(S_{0 ...i_k ...0})^2 \
        \nonumber\\
&&\ \ \ \ \ \ \ +\sum_{k,l=1}^n \sum_{i_k, i_l=1}^3(S_{0 ...i_k
...i_l ...0})^2 - ...
        \nonumber\\
&&\ \ \ \ \ \ \ +\ (-1)^n \sum_{i_1,...,i_n=1}^3(S_{i_1 ...
i_n})^2\bigg\} .
\end{eqnarray}

\noindent Here, we show this scalar to be useful for understanding
state purity and entanglement properties of multi-photon systems.
We note immediately that these satisfy the fundamental requirement
of entanglement measures that they be invariant under local
unitary transformations of SU(2), since these are a subgroup of
the O$_0(1,3)$ transformations of SLOCC under which they are
invariant, and since no renormalization is required after their
action. We see that the quantum state purity for a general
$n$-photon state can be written simply in terms of the
multi-photon Stokes parameters as
\begin{equation}
{1\over 2^n}\sum_{i_1, ..., i_n=0}^3S^2_{i_1, ...,
i_n}=\mathrm{Tr} {\rho^2}\ ,
\end{equation}
\noindent which is seen to be the corresponding Euclidean norm of
the multi-photon Stokes tensor. More importantly, we see that the
Stokes scalar can be expressed in terms of density matrices as
\begin{equation}
S^2_{(n)}=\mathrm{Tr} (\rho_{12... n}\, \tilde{\rho}_{12... n})\ ,
\end{equation}
\noindent where $\tilde{\rho}_{1... n}=(\sigma_2\otimes
\sigma_2\otimes ... \otimes\sigma_2)\rho_{1...
n}^*(\sigma_2\otimes \sigma_2\otimes ... \otimes\sigma_2)$ is the
``spin-flipped" density matrix. As we will show, $S_{(2)}^2$
captures the entanglement of important classes of multiple-photon
states. This relation makes the Stokes scalar of exceptional
interest as a candidate entanglement measure since, as shown
recently, functionals of the form $\mathrm{Tr}(\rho_a\rho_b)$ are
directly estimable through the visibility of interference arising
in an appropriate quantum network \cite{Ekert}, in addition to
being indirectly measurable via the quantum tomography approach
\cite{2}. We note that $S^2_{(n)}$ is of the required form, for
example $\rho_a=\rho_{12}$ and $\rho_b=\tilde{\rho}_{12}$ in the
case $n=2$.

The connection between our invariant and the basic Stokes
parameters in the $n=1$ case is simple. In that case, the Stokes
parameters form a vector of elements
$S_\mu=\mathrm{Tr}(\rho\sigma_\mu), \ \mu = 0, 1, 2, 3,$ and
$S^2_{(1)}=S_0^2-S_1^2-S_2^2-S_3^2$, similarly to the classical
case. In this case, the only relevant quantity is $S^2_{(1)}$. One
can relate the single-photon state purity to the Stokes scalar as
follows:
\begin{equation}\label{Pol1}
 \mathrm{Tr}(\rho^2)=1-S_{(1)}^2\ ,
\end{equation}
while $P^2= S_1^2+S_2^2+S_3^2=1-2S_{(1)}^2$ is the well-known
degree of photon polarization. Equivalently, we see that
$S^2_{(1)}$ is the single-photon linearized entropy,
\begin{equation}
S^2_{(1)}=1-\mathrm{Tr}\rho^2\
\end{equation}
(see, for example, \cite{Tetal}). $S^2_{(1)}$ allows us to
understand the effect of O$_o(1,3)$ transformations on state
purity. Under ordinary polarization rotations, $S_0$ itself
remains unchanged, so the purity is unchanged. However, the
hyperbolic polarization rotations filter the beam in a
basis-dependent way, reducing the quantum ensemble and diminishing
the intensity to $S'_0<1$. We notate the Stokes vector
transformation under an element of O$_0(1,3)$ as $S_\mu\rightarrow
S'_\mu$. Recalling that it is therefore necessary to renormalize
the state, $S'_\mu\rightarrow S''_\mu=S'_\mu /S'_0$, we see that
these filtering transformations effectively increase the Stokes
scalar in the single-photon case: $S''^2_{(1)}>S^2_{(1)}$, since
$S''^2_{(1)}\equiv S'^2_{(1)}/S'^2_0$ with $S'^2_{(1)}=S^2_{(1)}$
due to the invariance of $S^2_{(1)}$ under O$_0(1,3)$. In this
way, these filtering transformations are seen to decrease the
purity and increase the linearized entropy of single-photon
polarization states, just as it does in the classical case.

\section{The case $n=2$}

Our central interest here, however, is that of two or more
entangled photons. The generalized Stokes parameters that
characterize the two-photon polarization quantum ensemble are
\begin{equation}\label{Smunu}
  S_{\mu\nu}=\mathrm{Tr}(\rho\ \sigma_\mu\otimes\sigma_\nu)\ ,
\end{equation}
where $\mu , \nu= 0, 1, 2, 3$ \cite{1,2,3}. This collection of
Stokes parameters has 16 elements, which are systematically
measurable, as is done in quantum state tomography to determine
the density matrix, and capture all the polarization correlations
present in a photon pair as well as single-photon polarization and
beam intensity information \cite{3}. Consider the scalar invariant
for the case $n=2$,
\begin{equation}\label{S12}
S_{(2)}^2={1\over
4}\bigg\{(S_{00})^2-\sum_{i=1}^3(S_{i0})^2-\sum_{j=1}^3(S_{0j})^2
    \nonumber\\
\ +\ \sum_{i=1}^3\sum_{j=1}^3(S_{ij})^2\bigg\}\ .
\end{equation}
Note that the second and third terms of the r.h.s. of Eq.
(\ref{S12}) pertain only to the one-photon subsystems, being the
squares of the polarizations of the individual particles 1 and 2,
$P^2_1=S_{10}^2+S_{20}^2+S_{30}^2$ and
$P^2_2=S_{01}^2+S_{02}^2+S_{03}^2$, while the final term refers
only to the two-photon composite system \cite{17}. Again, the
two-photon state purity can be simply related to the Stokes scalar
$S_{(2)}^2$:

\begin{equation}\label{Rho12}
\mathrm{Tr}(\rho^2)=\bar{P^2}+S^2_{(2)} ,
\end{equation}

\noindent where $\bar{P^2}\equiv{1\over 2}(P_1^2+P_2^2)$ is the
average of the squares of the single-photon polarizations. More
important, however, is the fact that $S^2_{(2)}$ can be seen to be
a measure of entanglement for two-photon pure states. In fact, Eq.
(9) and some algebra show that in the case of pure states this
scalar coincides with the concurrence squared,  expressed in terms
of two-photon Stokes parameters ({\it cf.} \cite{3}) -- that is,
the tangle, $\tau$ \cite{W98}:
\begin{equation}\label{Tangle}
\tau =S_{(2)}^2,
\end{equation}
so the entanglement of formation is seen to be $h\Big({1\over
2}\Big[ 1+\sqrt{1-S_{(2)}^2}\ \Big]\Big)$, where $h(x)\equiv -x
\log_2\ x - (1-x) \log_2\ (1-x)$. However, the valuable property
of $S^2_{(2)}$, beyond its being equal the tangle for two-photon
pure states, appears in its application to {\it mixed} states,
which we consider next, where it is not equal to the square of the
concurrence.

Consider now the class of mixed states that describe two photons
of a three-photon system in a pure quantum state. For two-photon
mixed states, $S^2_{(2)}$ is different from the tangle, which is
not well-defined for mixed states. For this important class of
states, we find $S^2_{(2)}$ to be a specific sum of entanglement
measures over the pertinent subsystem and the larger,
three-particle system. To see this, recall that any three-photon
state can be written

\begin{equation}
|\Psi\rangle=\sum_{ijk}a_{ijk}|i\rangle_A |j\rangle_B |k\rangle_C\
.
\end{equation}

\noindent Examining the relationship of the entanglement of photon
A with a pair of photons B and C, we have from Eq. (5) that

\begin{equation}
S^2_{AB}+S^2_{AC}=C^2_{A(BC)}\ ,
\end{equation}

\noindent where $C^2_{A(BC)}$ is the concurrence calculated for a
bipartite decomposition of ABC into subsystem A and (composite)
subsystem BC \cite{CKW00}, and where $S_{AB}^2$ and $S_{AC}^2$ are
the values of $S_{(2)}^2$ for the two-photon subsystems AB and AC.
Furthermore, since $C^2_{A(BC)}=C^2_{AB}+C^2_{AC}+\tau_{ABC}$,
where $\tau_{ABC}$ is a three-particle entanglement measure (the
``three-tangle") \cite{CKW00}, we have that

\begin{equation}
S^2_{AB}+S^2_{AC}=C^2_{AB}+C^2_{AC}+\tau_{ABC}\ .
\end{equation}

\noindent Thus, we have that

\begin{equation}
\tau_{ABC} = (S^2_{AB}-C^2_{AB})+(S^2_{AC}-C^2_{AC})\ ,
\end{equation}

\noindent which shows that the sum of the $S_{(2)}^2$ values for
the two-photon subsystems AB and AC captures the contribution to
the total three-particle state entanglement encoded in these
two-particle subsystems, as well as their own internal
two-particle entanglements as measured by the concurrence. Similar
expressions are obtained when one begins with the other two
bipartite decompositions of ABC. By jointly considering the
resulting expressions, one finds that

\begin{equation}
S_{AB}^2=C_{AB}^2+\tau_{ABC}/ 2\ ,
\end{equation}

\noindent and similarly for $S_{AC}^2$ and $S_{BC}^2$.

Since the entanglement of formation is a monotonically increasing
function of both concurrence and three-tangle, it is a
monotonically increasing function of $S_{(2)}^2$ as well. We see
from Eq. (16) that $S^2_{(2)}$ captures its contribution to the
three-particle entanglement as well as the two-particle
entanglement of the corresponding subsystem. Eq. (16) shows it to
be an entanglement measure for three-photon pure states that
includes entanglement not present in the concurrence of its
two-particle subsystems. We also see from Eq. (15) that, though
the photon-pair contributions to the total entanglement may differ
as a result of their own internal entanglements, each of the
pairings AB and AC can be viewed as also containing the
three-tangle of the three-photon state $|\Psi\rangle$. This result
is analogous to what one finds for the entanglement of
single-qubit subsystems of a two-qubit system in a pure state,
where the reduced states of the two subsystems encode the tangle
of the overall system.

\section{Filtering operations}

Recall that, in order to be properly interpreted physically after
filtering, the n-photon Stokes parameters $\{S_{i_1...i_n}\}$ must
be renormalized, $S'_{i_1...i_n}\rightarrow
S''_{i_1...i_n}/S_{0...0}$. After a local filtering operation one
has $S''^2_{(n)}= S'^2_{(n)}/S^2_{0...0}$. The value of
$S''^2_{(2)}$ thus monotonically increases with filtering. For the
case $n=2$ this has a clear meaning in terms of entanglement. In
that case, $S'_{\mu\nu}\rightarrow S_{\mu\nu}''$: after a
filtering operation one has $S''^2_{(2)}=S^2_{(2)}/S'^2_{00}$. The
invariance of the scalar under O$_0(1,3)$ transformations
$S_{\mu\nu}\rightarrow S_{\mu\nu}'$, that is
$S'^2_{(2)}=S_{(2)}^2$, means that the effect of these
transformations on the invariant is entirely captured by the total
attenuation. The filtering operation thus results in an increase
of entanglement, since $S'_{00}<1$, by virtue of Equations (11)
and (16). Such local operations can be implemented, for example,
by dichroic optical fiber, where they are associated with
polarization-dependent losses (see, for example \cite{Gisin}).

The Stokes scalar $S^2_{(2)}$ allows us to identify and quantify
the beneficial effect of the SLOCC filtering operations on
entanglement for the classes considered above. Thus, we see how
the attenuating transformations [O$_0(1,3)\setminus$ SO(3)]
together with the quantum ensemble renormalization they engender,
correspond to entanglement purifying transformations (see also
\cite{G96}, \cite{HHH} and, in particular \cite{Vetal}). These
operations affect the invariant $S^2_{(n)}$ in exactly the same
way for general values of $n$.

\section{Conclusion}

In summary, we have introduced a group-invariant Stokes scalar for
studying $n$-photon entangled polarization states. In the case of
two-photon pure states, this invariant is equal to the tangle. In
the case of three-photon pure states, it measures entanglement in
the total system through its photon-pair subsystems, which are in
general described by mixed states. These results allow us to
identify sets of optical elements that give rise to
polarization-dependent filtering, such as dichroic optical fibers,
as entanglement purifiers, and to quantify their
entanglement-increasing effect on two-photon and three-photon
states. Such local transformations have a similar effect on the
invariant in the case of photon numbers $n$ greater than 3.

Because it satisfies the necessary condition of being invariant
under local unitary transformations for general values of $n$, and
since it has a clear connection to the accepted entanglement
measures of tangle and three-tangle characterizing few photon
states, this invariant can be considered a good candidate to
measure entanglement for pure states of $n$ photons. In addition,
it can be directly estimated, at least in principle, via a
suitable quantum network. Finally, unlike the proposed $n$-tangle
measure \cite{CW} for uniquely $n$-particle entanglement, which is
ill-defined for odd values of $n$, this invariant is well-defined
for all finite values of $n$.

This work was supported by the DARPA QuIST program, the National
Science Foundation (NSF), the Center for Subsurface Sensing and
Imaging Systems (CenSSIS, an NSF Engineering Research Center), and
the David and Lucile Packard Foundation.

\end{document}